\documentclass[twocolumn,preprintnumbers,amsmath,amssymb,superscriptaddress,nofootinbib,longbibliography]{revtex4-1}

\usepackage{graphicx}
\usepackage{bm}
\usepackage{dsfont}
\usepackage[usenames,dvipsnames]{xcolor}
\usepackage{pstricks}
\usepackage[tight]{subfigure}
\usepackage{verbatim}
\usepackage{units}
\usepackage{multirow}
\usepackage{enumitem}
\usepackage{mathrsfs}
\usepackage{leftidx}
\usepackage{xspace}
\usepackage{braket}
\usepackage{bbm}
\usepackage{cancel}
\usepackage{amsfonts,amssymb,amsmath}
\usepackage[normalem]{ulem}

\usepackage{fontawesome}

\usepackage[a4paper]{hyperref}
\hypersetup{colorlinks=true,linktoc=all,linkcolor=blue,breaklinks=true,citecolor=blue,urlcolor=blue}
\usepackage[english]{babel}
\newcommand{\definition}{:=}

\newcommand{\kB}{{k}_\text{B}}
\newcommand{\Tr}[2]{\text{Tr}_{#1}\left\{ #2\right\}}

\begin{document}

\title{Fluctuation-dissipation bounds for time-dependently driven conductors}

\author{Ludovico Tesser}
\affiliation{Department of Microtechnology and Nanoscience (MC2), Chalmers University of Technology, S-412 96 G\"oteborg, Sweden}

\author{Jos\'e Balduque}
\affiliation{Departamento de F\'isica Te\'orica de la Materia Condensada, Universidad Aut\'onoma de Madrid, 28049 Madrid, Spain\looseness=-1}
\affiliation{Condensed Matter Physics Center (IFIMAC), Universidad Aut\'onoma de Madrid, 28049 Madrid, Spain\looseness=-1}

\author{Janine Splettstoesser}
\affiliation{Department of Microtechnology and Nanoscience (MC2), Chalmers University of Technology, S-412 96 G\"oteborg, Sweden}
	
\date{\today}

\begin{abstract}
We analyze the noise in a multi-terminal multi-channel conductor under arbitrary time-dependent driving and subject to---possibly large---static potential and temperature biases. We show that the full out-of-equilibrium zero-frequency noise 
is constrained by a \textit{fluctuation-dissipation bound}.
It consists of an upper bound expressed in terms of  weighted current components of the separate Floquet bands arising from the time-dependent driving.  In the limit of large static temperature bias, it has an intuitive interpretation in terms of the dissipated powers due to the static potential bias and due to the time-dependent driving. 
Furthermore, we show the existence of a second bound that relies on the specific shape of the electron distribution resulting from the driving, which is often even tighter than the fluctuation-dissipation bound.
We show the implications of our bounds at the simple, but experimentally relevant example of a two-terminal conductor in the presence of an ac bias. 
\end{abstract}

\maketitle

\section{Introduction}

Precision plays an important role in small-scale systems, where the magnitude of fluctuations can be significant with respect to the desired average observable.
For example, thermoelectric nanoscale conductors can serve as engines~\cite{Benenti2017Jun,Whitney2018May,Cangemi2024Oct,Balduque2025Apr}, transforming tiny amounts of heat into electrical power at a few-particle level or vice versa by using electrical power as a resource for cooling~\cite{Giazotto2006Mar}.
In quantum transport, driven nanoscale conductors furthermore serve as precise current sources~\cite{Pekola2013Oct,Edlbauer2022Dec}.
Hence, knowledge about how the the power and current fluctuations behave is crucial~\cite{Blanter2000Sep,Kobayashi2021Sep}.
It is of particular interest to understand how the fluctuations are related to the desired \textit{average} output. At equilibrium or close to equilibrium, a relation between noise and average response is provided by fluctuation-dissipation theorems~\cite{Callen1951Jul, Green1954Mar, Kubo1957Jun,Esposito2009Dec}. But even out of equilibrium, similar relations have been found for systems with a weak tunnel coupling and subject to a voltage bias~\cite{Rogovin1974Jul,Levitov2004Sep,Andrieux2006Jan} or to time-dependent driving~\cite{Safi2014Jan,Riwar2021Jan}.
Also, far from equilibrium fluctuation-dissipation theorem-like relations can be established under stalling conditions~\cite{Altaner2016Oct,Shiraishi2022Jul} or using cumulant expansions of the full-counting statistics~\cite{Tobiska2005Dec,Forster2008Sep,Utsumi2010Mar}.

\begin{figure}[b]
    \centering
    \includegraphics[width=0.7\linewidth]{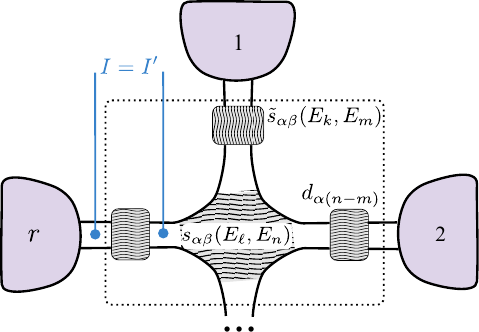}
    \caption{Sketch of the multi-terminal setup with contacts $\alpha=1,..,r$. Time-dependent driving is applied in the leads and in the central region (patterned). Measurements of the time-averaged current and of the zero-frequency noise,  when performed before or after the driven lead region (blue spots), yield the same result.  }
    \label{fig:setup}
\end{figure}

When releasing the constraints set by the specific out-of-equilibrium conditions, fluctuation-dissipation theorems are not available any longer. Nonetheless, important information on the minimum amount of fluctuations that occur in a process have been formulated in terms of fluctuation \textit{bounds}~\cite{Landi2024Apr}. These bounds have different purposes and origins: the so-called thermodynamic uncertainty relation~\cite{Barato2015Apr,Gingrich2016Mar} constrains the precision by the entropy production; the kinetic uncertainty relation constrains the precision in terms of the activity of a process~\cite{DiTerlizzi2018Dec}. 
Both of these bounds were originally developed for classical processes. The challenge of treating coherent quantum transport in conductors that are possibly strongly coupled to the contacts has been tackled in Refs.~\cite{Brandner2018Mar,Palmqvist2024Oct,Palmqvist2025Apr,Brandner2025Feb,Timpanaro2025Jan}, and also bounds on the full statistics have very recently been addressed~\cite{Brandner2025Jul}. Furthermore, for this coherent, strong coupling situation, recently, a fluctuation-dissipation bound was developed, constraining the amount of nonequilibrium noise by the nonequilibrium conditions (temperature and voltage bias) and by the desired current flow resulting from it~\cite{Tesser2024May}. However, noise bounds for coherent quantum transport under \textit{time-dependent driving}, have until now been limited to thermodynamic bounds in terms of entropy production and to linear response~\cite{Potanina2021Apr} or slow driving~\cite{Lu2022Mar}. Also, recently developed more generally valid bounds on the transition rates provide simple statements for the noise only in the regime of weak tunnel coupling~\cite{Tesser2025Apr}. 
Understanding limits on noise in the presence of arbitrary time-dependent driving \textit{and} temperature bias, is however important for the precision of cyclically operating heat engines as well as for time-dependently driven conductors in which accidental temperature differences arise due to the operation of the device, where they can strongly impact the precision.

In this paper, we address this shortage and extend the nonequilibrium fluctuation-dissipation bound of Ref.~\cite{Tesser2024May}, which is valid in the stationary regime, to systems with generic time-dependent driving on top of static out-of-equilibrium conditions set by temperature and potential differences. 
This includes both driving of the central conductor, for example by a modulation of gate voltages, as well as time-dependent driving applied to the contacts, such as time-dependent bias voltages or even time-dependent temperatures~\cite{Portugal2024Jun} which can be modeled by an effective time-dependent Hamiltonian~\cite{Luttinger1964Sep}. 
We establish a bound on the noise, more precisely on the zero-frequency charge-current autocorrelators, by comparing situations with and without voltage and temperature biases, in the presence of the driving.  
We show that the nonequilibrium fluctuations are bounded by a weighted sum over spectral current components. This fluctuation-dissipation bound for time-dependent systems, in the presence of considerable temperature differences, equals the power provided by or absorbed by the drive and the power dissipated due to the stationary nonequilibrium conditions. 
We furthermore develop an additional, so-called intersection bound, which is often tighter than the fluctuation-dissipation bound but requires knowledge of the full electronic distribution functions modified by the time-dependent driving.

In order to set up these bounds, we use scattering theory, thereby fully treating quantum coherences and strong coupling, which is in contrast to typical stochastic approaches~\cite{Esposito2009Dec,Landi2024Apr}. This comes at the cost of treating Coulomb interactions only up to the mean-field level. Note, however, that the fact that our bounds hold for scattering matrices for an \textit{arbitrary} time-dependent driving has an important implication: our theory fully includes the treatment of possible screening potentials due to the time-dependent driving that affect currents and noise~\cite{Christen1996Sep,Sanchez2013Jan,Meair2013Jan,Texier2018Feb,Dashti2021Dec}.

The remainder of this paper is organized as follows. We introduce the model for the driven multi-terminal, multi-channel conductor in Sec.~\ref{sec:model}. In Sec.~\ref{sec:approach}, we also show how relevant observables, namely charge currents and charge-current noise as well as energy- and heat-currents revealing the role of the dissipated or provided power, are obtained from Floquet scattering theory for time-dependently driven conductors. We then derive the out-of-equilibrium dissipation bound for time-dependently driven systems in Sec.~\ref{sec:FDB} and show its intuitive interpretation in the limit of large temperature bias in Sec.~\ref{subsec:large_Delta_T}. The alternative intersection bound in terms of effective distribution functions is presented in Sec.~\ref{subsec:intersection}. In Sec.~\ref{sec:plots}, these bounds are demonstrated in the example of an ac-biased two-terminal conductor in the presence of a temperature difference. Detailed derivations are provided in the Appendices.

\section{Model and Approach}\label{sec:approach}

\subsection{Multi-terminal setup}\label{sec:model}

We study a multi-terminal, multi-channel conductor, subject to stationary out-of-equilibrium conditions due to different temperatures and electrochemical potentials in the contacts and subject to arbitrary time-dependent driving. We describe the setup, sketched in Fig.~\ref{fig:setup}, by Floquet scattering theory~\cite{Moskalets2011Sep}, see also Refs~\cite{Tien1963Jan,Shirley1965May}.

The contacts of this multi-terminal setup are labeled by greek letters $\alpha,\beta,\gamma=1,...,r$. Each contact $\alpha$ is described by a macroscopic distribution function, characterized by Fermi functions with a given temperature $T_\alpha=\bar{T}+\Delta T_\alpha$ compared to an equilibrium temperature $\bar{T}$ and electrochemical potential $\mu_\alpha=\bar{\mu}+qV^\mathrm{dc}_\alpha$ compared to an equilibrium potential $\bar{\mu}$. 
Here, we introduced the charge of the quasiparticle excitations $q$, which is typically going to be the electron charge. Excitations leaving the contact and impinging on the scattering region via channel $n=1,...,N_\alpha$, are characterized by creation and annihilation operators, $\hat{a}_{\alpha n}^\dagger(E)$ and $\hat{a}_{\alpha n}(E)$, fulfilling
\begin{equation}
    \langle \hat{a}^\dagger_{\alpha n}(E)\hat{a}_{\beta m}(E')\rangle=\delta_{\alpha \beta }\delta_{nm}\delta(E-E')f_\alpha(E)
\end{equation}
with channel index $n$ counting the $N_\alpha$ channels in contact $\alpha$. Here, $f_\alpha(E)$ denotes the Fermi function and we will later use the definition $f_\alpha^-(E) \definition 1-f_\alpha(E)$. 
These excitations get scattered in the conductor, while picking up an integer amount of Floquet quanta due to the time-dependent driving. 
By contrast, the excitations leaving the scatterer and impinging on the contacts are characterized by creation and annihilation operators, $\hat{b}_{\alpha n}^\dagger(E)$ and $\hat{b}_{\alpha n}(E)$, fulfilling
\begin{eqnarray}\label{eq:stilde}
\langle \hat{b}_{\alpha n}^\dagger (E)\hat{b}_{\alpha n} (E')\rangle =  \tilde{s}^*_{\alpha n,\beta m}(E,E_k)\tilde{s}_{\alpha n,\beta m}(E',E_\ell')\nonumber\\
 \times f_\beta(E_{k})\ \delta(E-E'-(\ell-k)\hbar\Omega)
\end{eqnarray}
where the sum over \textit{additional} indices appearing on the right-hand side is from here on always implicit, if not otherwise indicated. 
Here, we have introduced the Floquet scattering matrix elements $\tilde{s}_{\alpha n,\beta m}(E,E_k)$, which provide the amplitude for an excitation incoming from channel $m$ in contact $\beta$ at energy $E_k=E+k\hbar\Omega$ to be scattered into channel $n$ of contact $\alpha$, while exchanging $-k$ Floquet quanta $\hbar\Omega$. 
This exchange of Floquet quanta in the scattering process arises from the time-dependent driving, which we here assume to result from the driving of any set of parameters $\left\{X(t)\right\}$ that can be decomposed in a Fourier series with frequency $\Omega$, namely $X(t)=\sum_n e^{-in\Omega t}X_n$. 
The scattering matrix fulfills a unitarity condition~\cite{Moskalets2011Sep}, see also Appendix~\ref{app:properties}. 

In order to model different types of experimentally relevant settings, it can be useful to decompose the scattering matrix into parts describing the back-scattering free evolution due to driving in the lead regions and scattering between different contacts under the influence of driving in the central region, see Fig.~\ref{fig:setup},
\begin{equation}
\tilde{s}_{\alpha n,\beta m}(E,E_{k})  \equiv    d_{\alpha\ell}s_{\alpha n,\beta m}(E_{-\ell},E_{k+p})c_{\beta p}  .
\end{equation}
 Here, we introduced the Floquet scattering matrix of the central region only, $s_{\alpha n,\beta m}(E_{-\ell},E_{k+p})$, as well as the Floquet coefficients due to the ac-driving in the leads, $d_{\alpha\ell}$, $c_{\beta p}$ which are defined as
\begin{subequations}\label{eq:Floquet_coefficients}
\begin{eqnarray}
d_{\alpha\ell} & = & \int_0^\mathcal{T}\frac{dt}{\mathcal{T}}e^{-i\phi_\alpha(t)}e^{i\ell\Omega t} \\
\phi_\alpha(t) & = & \frac{q}{\hbar}\int_0^t dt'V_\alpha^\mathrm{ac}(t') \\ V_\alpha^\mathrm{ac}(t) & = & \sum_nV_{\alpha n}e^{-in\Omega t}
\end{eqnarray}
\end{subequations}
and equivalently for $c_{\beta p}$ for a generally different ac-driving potential. Here, $\mathcal{T}=2\pi/\Omega$ is the period of the drive. Note that introducing the lead driving with potential $V_\alpha^\mathrm{ac}(t)$ is also a convenient way to model the ac part of a time-dependent bias voltage in a contact~\cite{Moskalets2011Sep}. The scattering matrix $s_{\alpha n,\beta m}(E_{-\ell},E_{k+p})$ can in principle contain any type of effects due to time-dependent driving, including screening-induced time-dependent potentials.

\subsection{Currents and fluctuations}\label{sec:observables}

We are interested in the time-averaged charge current and its zero-frequency noise. To evaluate these observables, we start from the time-dependent current operator
\begin{eqnarray}\label{eq:I_op}
    \hat{I}_\alpha(t) & = & \frac{q}{h}\int dE\ dE'\ e^{i(E-E')t/\hbar}\\
    && \times\left[\hat{b}^\dagger_{\alpha n}(E)\hat{b}_{\alpha n}(E')-\hat{a}^\dagger_{\alpha n}(E)\hat{a}_{\alpha n}(E')\right]\ .\nonumber
\end{eqnarray}
Evaluating the expectation value, $I_\alpha(t)=\langle\hat{I}_\alpha(t)\rangle$, and its time average, $I_\alpha=\int_0^\mathcal{T} I_\alpha(t)dt/\mathcal{T}$, namely the dc component of the current, we find
\begin{eqnarray}\label{eq:Iaverage}
    I_\alpha
    &=&q\int \frac{dE}{h} \\
    &&\left( \Tr{}{\tilde{t}_{\alpha\beta}^\dagger(E,E_k)\tilde{t}_{\alpha\beta}(E,E_{k})}f_{\beta}(E_k) - N_\alpha f_\alpha(E)\right).\nonumber
\end{eqnarray}
Here, we have introduced sub-matrices $\tilde{t}_{\alpha\beta}(E,E_k)$ of the scattering matrix with elements $\left[\tilde{t}_{\alpha\beta}(E,E_k)\right]_{nm}=\tilde{s}_{\alpha n,\beta m}(E,E_{k})$ to keep the notation compact; the trace in Eq.~\eqref{eq:Iaverage} is hence taken over all channels $n=1,...,N_\alpha$, while additional (implicit) sums are performed over the indices $\beta$ and $k$. 
Note that in the leads no backscattering is induced due to the driving. Therefore, a measurement of the time-averaged current (as well as of the zero-frequency charge-current noise) before or after the driven lead region, indicated by blue spots in Fig.~\ref{fig:setup}, yields identical results due to current conservation. Starting from the current, we calculate the linear conductances $G_{\alpha\beta}\definition \partial I_\alpha/\partial V_\beta|_{\{\mu_\beta = \mu_\alpha, T_\beta = T_\alpha\}}$ for $\alpha\neq\beta$,
\begin{eqnarray}\label{eq:lin_conductance}    
    G_{\alpha\beta} &
   = & \frac{q^2}{h}\int dE\ \Tr{}{\tilde{t}^\dagger_{\alpha\beta}(E_k,E_{})\tilde{t}_{\alpha\beta}(E_k,E_{})}\nonumber\\
   && \times\frac{1}{\kB T_\alpha}f_\alpha(E)f^-_\alpha(E).
\end{eqnarray}
Note that this is the conductance in the presence of full time-dependent driving. 

In analogy to the charge current, the energy current \textit{flowing into contact $\alpha$} in the driven system is given by~\cite{Butcher1990Jun,Moskalets2011Sep}
\begin{eqnarray}
 I_\alpha^E & = &  \int dE\ \frac{E}{h}\Tr{}{\tilde{t}^\dagger_{\alpha\beta}(E,E_{k})\tilde{t}_{\alpha\beta}(E,E_{k})}\nonumber\\
   && \times \left(f_\beta(E_{k})- f_\alpha(E)\right)\ .\label{eq:Ecurrent}
\end{eqnarray}
The energy current is the starting point to calculate heat currents and dissipated power, see Sec.~\ref{subsec:large_Delta_T} and Appendix~\ref{app:power}, which plays an important role in the fluctuation-dissipation bound we develop.

Indeed, the quantity of central interest here is the noise. Concretely, we consider the zero-frequency auto-correlations in one of the contacts. In the following Sec.~\ref{sec:FDB}, we always assume that the noise is measured in the hottest contact. Nonetheless, we here start from the definition for the zero-frequency noise of $\hat{I}_\alpha$ in any contact $\alpha$,
\begin{equation}\label{eq:def_correlator}
    \mathcal{S}_{\alpha\alpha} \definition \int_0^\mathcal{T}\frac{dt}{\mathcal{T}}\int d\tau\langle \delta \hat{I}_\alpha(t+\tau)\delta\hat{I}_\alpha(t)\rangle .
\end{equation}
with $\delta \hat{I}_\alpha(t) \definition \hat{I}_\alpha(t) - I_\alpha$.
Using the definition of the current operator, given in Eq.~\eqref{eq:I_op}, and plugging in the full Floquet scattering matrix~\eqref{eq:stilde}, we find
\begin{equation}\label{eq:fullnoise}
\begin{split}    
& \mathcal{S}_{\alpha\alpha} = \frac{q^2}{h}\int dE f_\beta(E_{k_1})f^{-}_\gamma(E_{k_2})\\ &\times\mathrm{Tr}\left\{\left(\tilde{t}^\dagger_{\alpha\beta}(E,E_{k_2})\tilde{t}_{\alpha\gamma}(E,E_{k_2})-\delta_{\alpha\beta}\delta_{\alpha\gamma}\delta_{k_1k_2}\delta_{k_1\ell_2}\delta_{k_2\ell_1}\right)\right.\\
& \times\left.\left(\tilde{t}^\dagger_{\alpha\gamma}(E_{k_2-\ell_1},E_{k_2})\tilde{t}_{\alpha\beta}(E_{k_1-\ell_2},E_{k_1})-\delta_{\alpha\beta}\delta_{\alpha\gamma}\delta_{k_1k_2}\right)\right\}\ .
 \end{split}
 \end{equation}
We recall that all additional (Floquet and contact) indices occurring on the right hand side are summed over. We evaluate this expression starting from terms that are of zeroth, second and fourth order in the scattering matrices, $\mathcal{S}_{\alpha\alpha} = \mathcal{S}_{\alpha\alpha}^{(0)} + \mathcal{S}_{\alpha\alpha}^{(2)} +\mathcal{S}_{\alpha\alpha}^{(4)}  $ with 
\begin{subequations}
\label{eq:S_contributions}    
\begin{eqnarray}
    \mathcal{S}_{\alpha\alpha}^{(0)} & \definition &  \frac{q^2}{h} \int dE  
    f_\alpha(E)f^-_\alpha(E)N_\alpha\ \label{eq:S0_contribution}\\
 \mathcal{S}_{\alpha\alpha}^{(2)}  & \definition & -2\frac{q^2}{h} \int dE 
   \Tr{}{\tilde{t}^\dagger_{\alpha\alpha }(E,E_{k})\tilde{t}_{\alpha\alpha}(E,E_{k})} \nonumber\\
   &&\times f_\alpha(E_{k})f^-_\alpha(E_{k}),\label{eq:S2_contribution}
\end{eqnarray}
as well as 
\begin{eqnarray}
      {S}_{\alpha\alpha}^{(4)} &\definition& \frac{q^2}{h}\int dE \mathrm{Tr}\left\{{\tilde{t}}_{\alpha\beta}(E_{\ell},E_{k})\tilde{t}_{\alpha\beta}^\dagger (E,E_{k})\right. \label{eq:S4_contribution}\\
      &&\left.\times\tilde{t}_{\alpha\gamma}(E,E_{p})\tilde{t}_{\alpha\gamma}^\dagger (E_{\ell},E_{p}) \right\}f_\beta(E_{k}) f_\gamma^-(E_{p})\ .\nonumber
\end{eqnarray}
\end{subequations}
Below we will use these compact expressions to derive the out-of-equilibrium fluctuation-dissipation bound for time-dependently driven systems, as well as an alternative bound we refer to as intersection bound, in Sec.~\ref{sec:FDB} and Appendix~\ref{app:inequalities}.

\section{Fluctuation-dissipation bounds}\label{sec:FDB}

In the following we derive general bounds on the noise, valid for any realization of the scattering matrix, any stationary bias, and any type of periodic time-dependent driving. We will express these bounds in terms of an \textit{excess} noise, namely comparing the noise of the full nonequilibrium system to an expression containing linear conductances; in equilibrium, this expression equals the thermal noise.

\subsection{Fluctuation-dissipation bound for time-dependently driven systems}\label{subsec:FDB}

\begin{figure}
    \centering
\includegraphics[width=0.9\linewidth]{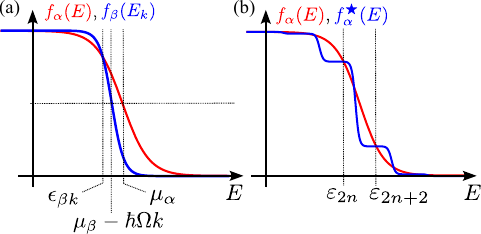}
    \caption{(a) Crossing of the Fermi function characterizing the hot contact with the Fermi function characterizing the cold contact, shifted by $k$ Floquet quanta. (b) Crossing of the occupation of the hot contact with the occupation stemming from the other contact(s) modified by the driving (here for an example with ac-driving in the contacts and full transmission $D=1$).}
    \label{fig:crossing}
\end{figure}

As a first step towards the derivation of the fluctuation-dissipation bound in the presence of time-dependent driving, we establish how the noise under full nonequilibrium conditions, namely in the presence of voltage and temperature biases \textit{and} in the presence of an arbitrary driving, relates to the linear conductances found when all \textit{static} electrochemical potentials and temperatures are the same.
We therefore first rewrite the term $ \mathcal{S}_{\alpha\alpha}^{(2)}$ given in Eq.~\eqref{eq:S2_contribution} in terms of a contribution that contains the linear conductance of Eq.~\eqref{eq:lin_conductance},
\begin{eqnarray}
    \mathcal{S}_{\alpha\alpha}^{(2)}   & = & 2 k_\mathrm{B}T_\alpha \sum_{\beta\neq\alpha}G_{\alpha\beta}+\nonumber\\
    && -2\frac{q^2}{h} \int dE 
   N_\alpha f_\alpha(E)f^-_\alpha(E),\label{eq:S23_contribution}
\end{eqnarray}
together with an additional term in the second row that will partially cancel with $\mathcal{S}_{\alpha\alpha}^{(0)} $, see Eq.~\eqref{eq:S0_contribution}. 
Highlighting the linear conductance in the presence of driving has the advantage that it provides the opportunity for connections to the thermal noise, since it appears in the equilibrium fluctuation-dissipation theorem.
Furthermore, rewriting the remaining term that is quartic in the scattering amplitudes, $\mathcal{S}_{\alpha\alpha}^{(4)}$ given in~\eqref{eq:S4_contribution}, and using the Cauchy-Schwarz inequality as explained in Appendix~\ref{app:inequalities}, we find a constraint on the full noise with respect to the linear conductances 
\begin{eqnarray}
  &&  \mathcal{S}_{\alpha\alpha} -  2\kB T_\alpha\sum_{\beta\neq\alpha}G_{\alpha\beta} \leq \nonumber\\   
   &&\leq \frac{q^2}{h} \int dE  \Tr{}{\tilde{t}_{\alpha\beta}
    (E,E_{k})\tilde{t}^\dagger_{\alpha\beta}(E,E_{k})}\nonumber\\ &&\hspace{1cm}\times\left(1-2f_\alpha(E)\right)\left(f_\beta(E_{k})-f_\alpha(E)\right)\ .\label{eq:start_bound}
\end{eqnarray}
The expression on the right-hand side is similar to the current, apart from the factor $1-2f_\alpha(E)$ in the integrand. Since we know that $1-2f_\alpha(E)$ is an increasing function in energy, we can bound the right-hand side of the inequality~\eqref{eq:start_bound} by opportunely replacing $1-2f_\alpha(E)$ with a convenient constant value, which can then be taken out of the energy integral. To make this replacement, we now choose contact $\alpha$, in which the current is measured, to be the \textit{hottest} one, i.e. $T_\alpha > T_\beta$ for all $\beta\neq\alpha$.  We then identify the energy at which the Fermi functions $f_\beta(E_{k}),f_\alpha(E)$ cross for each value of $k$. This crossing energy, indicated in Fig.~\ref{fig:crossing}{(a)}, is 
\begin{equation}\label{eq:crossing}
\epsilon_{\beta k}\definition \frac{T_\alpha(\mu_\beta-\hbar\Omega k)-T_\beta\mu_\alpha}{T_\alpha-T_\beta}\ .
\end{equation}
At energies smaller than the crossing energy, we have $f_\beta(E_{k})-f_\alpha(E)>0$ as well as $1-2f_\alpha(E)<1-2f_\alpha(\epsilon_{\beta k})$, while at energies larger than the crossing energy, we have $f_\beta(E_{k})-f_\alpha(E)<0$ as well as $1-2f_\alpha(E)>1-2f_\alpha(\epsilon_{\beta k})$. We can therefore write the following constraint for the noise
\begin{eqnarray}
\mathcal{S}_{\alpha\alpha} - 2\kB T_\alpha\sum_{\beta\neq\alpha}G_{\alpha\beta}\leq 
    q\sum_{\beta,k}\left(1-2f_\alpha(\epsilon_{\beta k})\right)I_{\alpha\beta, k}\label{eq:fullbound}
\end{eqnarray}
in terms of the contact-resolved current components for each Floquet band
\begin{eqnarray}
    I_{\alpha\beta,k}&\definition&q\int \frac{dE}{h}\Tr{}{\tilde{t}_{\alpha\beta}
    (E,E_{k})\tilde{t}^\dagger_{\alpha\beta}(E,E_{k})}\nonumber\\ &&\times\left(f_\beta(E_{k})-f_\alpha(E)\right)\ .\label{eq:current_components}
\end{eqnarray}
The constraint~\eqref{eq:fullbound} is the central result of this paper. We refer to it as the fluctuation-dissipation bound for time-dependent driving (t-FDB). It shows that the full nonequilibrium zero-frequency noise compared to the linear response function in the absence of static biases is bounded by a weighted sum of spectral currents. In the absence of driving, it reduces to the stationary FDB presented in Ref.~\cite{Tesser2024May}.
Using the explicit expression for the crossing energy in Eq.~\eqref{eq:crossing}, the t-FDB~\eqref{eq:fullbound} can be rewritten as
\begin{eqnarray}
&& \mathcal{S}_{\alpha\alpha} - 2\kB T_\alpha\sum_{\beta\neq\alpha}G_{\alpha\beta}\nonumber\\
&& \hspace{0.5cm}\leq -q\sum_{\beta,k}   \tanh\left\{\frac{\mu_\alpha-\mu_\beta+\hbar\Omega k}{2\kB(T_\alpha-T_\beta)}\right\}I_{\alpha\beta, k}\ .\label{eq:fullbound_rewrite}
\end{eqnarray}
In the following, we provide an intuitive interpretation of this bound, which is particularly relevant in the limit of large temperatures biases. 

\subsection{Limit of large temperature bias}\label{subsec:large_Delta_T}

Large temperature biases are of special interest for, e.g., heat engines. At the same time, typical extensions of the equilibrium fluctuation-dissipation theorem fail at large temperature differences. 
When the nonequilibrium situation is established by large temperature biases, $\Delta T_{\alpha\beta}\equiv T_\alpha-T_\beta$ compared to the sum of potential biases $\Delta \mu_{\alpha\beta}\equiv \mu_\alpha-\mu_\beta$ and a relevant number of Floquet quanta, namely for $\kB\Delta T_{\alpha\beta}\gg \Delta\mu_{\alpha\beta}+\hbar\Omega k$, the weighting factor of the current components in Eq.~\eqref{eq:fullbound_rewrite} can be expanded. 
Importantly, this situation cannot be covered by any close-to-equilibrium FDT~\cite{Callen1951Jul,Green1954Mar,Kubo1957Jun} or by extensions to finite bias voltage~\cite{Rogovin1974Jul,Levitov2004Sep} or driving~\cite{Safi2014Jan,Riwar2021Jan} where an overall equilibrium temperature is assumed. 
By studying this limit, we thus complement previous results for the important case where a quantum conductor is subject not only to driving, but also to a large temperature differences. We find in this limit
 \begin{eqnarray}
 &&\mathcal{S}_{\alpha\alpha} - 2\kB T_\alpha\sum_{\beta\neq\alpha}G_{\alpha\beta}\leq \label{eq:largebiasbound}\\
&&\leq    -\frac{q}{2\kB}\left(\sum_{\beta,k}   \frac{\Delta\mu_{\alpha\beta}}{\Delta T_{\alpha\beta}}I_{\alpha\beta, k}  +\sum_{\beta,k}\hbar\Omega k \frac{I_{\alpha\beta, k}}{\Delta T_{\alpha\beta}}\right)\nonumber.
 \end{eqnarray}  
These two contributions are related to the power due to the potential bias and due to the driving, which, as shown in Appendix~\ref{app:power}, fulfill
\begin{eqnarray}
    \sum_{\alpha,\beta,k}   \Delta\mu_{\alpha\beta}I_{\alpha\beta, k} & = &2q\bar{\mathcal{P}}^\text{pot}\\
    \sum_{\alpha,\beta, k} \hbar\Omega k I_{\alpha\beta, k} & = & 2q\bar{\mathcal{P}}^\text{driv}\ .
\end{eqnarray}
Here, we define the symmetrized power $\bar{\mathcal{P}}^\mathrm{x} = \frac{1}{2}\left(\mathcal{P}^\text{x}+\mathcal{P}^\text{x,tr}\right)$ as the average between the power in the actual and in the time-reversed (tr) system. 
The expressions in~\eqref{eq:largebiasbound} hence correspond to the lead-resolved contributions $\bar{\mathcal{P}}^\mathrm{x}_{\alpha\beta}$ to these symmetrized powers,  $\bar{\mathcal{P}}^\mathrm{x}=\sum_{\alpha\beta}\bar{\mathcal{P}}^\mathrm{x}_{\alpha\beta}$ . 
With this, we write the bound in the limit of large temperature biases as
\begin{equation}
\mathcal{S}_{\alpha\alpha} - 2\kB T_\alpha\sum_{\beta\neq\alpha}G_{\alpha\beta}\leq 
    -q^2\sum_{\beta}\frac{\left(\bar{\mathcal{P}}^\text{pot}_{\alpha\beta}+\bar{\mathcal{P}}^\text{driv}_{\alpha\beta}\right)}{\kB\Delta T_{\alpha\beta}} \label{eq:largebiasbound_power}\ .
\end{equation}
In a system where time-reversal symmetry is not broken, the symmetrized functions $\bar{\mathcal{P}}^\text{x}$ equal the powers $\mathcal{P}^\text{x}=\mathcal{P}^\text{x,tr}$. 
This is, for example, the case for the two-terminal system discussed in Sec.~\ref{sec:plots}, where the driving is applied only to one of the leads and not to the energy-dependent central scattering region connecting different contacts. 
The bound~\eqref{eq:largebiasbound_power} hence shows that it depends on whether the power is dissipated ($\mathcal{P}<0$) or produced ($\mathcal{P}>0$) to which extent the out-of-equilibrium noise is allowed to be larger or constrained to be smaller than the equilibrium-like noise in the absence of temperature and voltage biases. This interpretation of the noise bounds in terms of dissipated or generated power is an asset of the t-FDB compared to the intersection bound presented in the next Sec.~\ref{subsec:intersection} (which instead turns out to often be a tighter bound). In the absence of driving, the bound~\eqref{eq:largebiasbound_power} reproduces the result obtained in Ref.~\cite{Tesser2024May}, where power is dissipated due to an applied stationary bias voltage or produced due to a temperature bias in coherent conductors with thermoelectric (energy-filtering) properties.

\subsection{Intersection bound for nonthermal distributions}\label{subsec:intersection}

The time-dependent driving applied to the leads or to the central conductor as well as the ``mixing" of occupations from different baths result in modified distributions entering a given contact.
Such modified distributions in systems where thermalization is hindered are also referred to as athermal~\cite{Aguilar2022Apr,Aguilar2024Sep} or nonthermal~\cite{Sanchez2019Nov,Deghi2020Jul,Tesser2023Apr} distributions.
Another way of setting up a noise bound, complementary to~\eqref{eq:fullbound_rewrite}, is by identifying the crossings of such nonthermal distribution functions. Therefore, we start from relation~\eqref{eq:start_bound} and define the modified distribution function
\begin{equation}\label{eq:fbigstar}
    f^\bigstar_\alpha(E) = \frac{1}{N_\alpha}\Tr{}{\tilde{t}_{\alpha\beta}
    (E,E_{k})\tilde{t}^\dagger_{\alpha\beta}(E,E_{k})}f_\beta(E_{k})\ .
\end{equation}
This modified distribution function can be either larger or smaller than the hot distribution $f_\alpha(E)$ in different energy intervals. This means that sign changes in the difference between them occur at given energies $\varepsilon_n$, with $n=1,...,M$. The $M$ crossing energies between the occupation functions, see Fig.~\ref{fig:crossing}{(b)} for an example, are found from $g(\varepsilon)\equiv f^\bigstar_\alpha(\varepsilon) -f_\alpha(\varepsilon)=0$ with $g'(\varepsilon)\neq 0$.
Note that, since $f_\alpha$ is the hottest distribution, there is a (low) energy below which $f_\alpha(E)< f_\alpha^\bigstar(E)$ and a (high) energy above which $f_\alpha(E)> f_\alpha^\bigstar(E)$. This guarantees the number of crossing points $M$ between $f_\alpha(E)$ and $f_\alpha^\bigstar(E)$ to be odd. With this, we find the bound 
\begin{align}
&   \mathcal{S}_{\alpha\alpha} -  2\kB T_\alpha\sum_{\beta\neq\alpha}G_{\alpha\beta} \leq q^2 \sum_{n=0}^{\frac{M-1}{2}}(1-2f_\alpha(\varepsilon_{2n+1})) \nonumber\\\label{eq:intersection_bound}
&\hspace{2cm}\times N_\alpha\int_{\varepsilon_{2n}}^{\varepsilon_{2n+2}} \frac{dE}{h} \left(f^\bigstar_\alpha(E)-f_\alpha(E)\right) ,
\end{align}
where we set the convention $\varepsilon_0=-\infty$ and $\varepsilon_{M+1}=\infty$. The second line of~\eqref{eq:intersection_bound} corresponds to the current contribution from the energy window $[\varepsilon_{2n},\varepsilon_{2n+2}]$. We refer to the constraint~\eqref{eq:intersection_bound} as the ``intersection bound". This bound can be expected to often be tighter than the t-FDB~\eqref{eq:fullbound}, since the intersection bound relies on estimates adapted separately for energy intervals limited by the identified crossing points. 
This comes at the cost of providing less physical insight than the t-FDB, since the position of the crossings is not generally known, but needs to be evaluated for each specific nonthermal distribution $f^\bigstar_\alpha(E)$. Therefore, also a direct connection of the bound to other transport quantities such as the total dissipated power, in analogy to~\eqref{eq:largebiasbound_power}, cannot be straightforwardly established.

Note that intersection bounds of this type can be found in different shapes, depending on the definition of the effective distribution function, namely depending on which part of the full scattering matrix is incorporated into the modified distribution function, see also Appendix~\ref{app:intersection}. 
We anticipate that these types of intersection bounds could also be extended to arbitrary nonthermal distributions, which are not necessarily generated by a time-dependent driving, in contact with one thermal ``hotter" one. Here,  hotter means that the distribution is smaller than the nonthermal one for $E\to-\infty$ and larger than the nonthermal one for $E\to\infty$.

\section{Two-terminal conductor driven by an ac bias voltage}\label{sec:plots}

\begin{figure}
    \centering
    \includegraphics[width=0.8\linewidth]{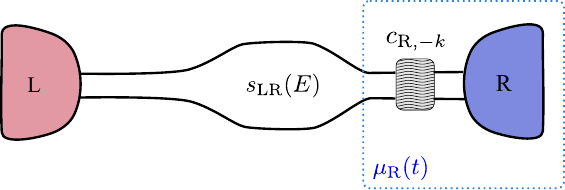}
    \caption{Sketch of the two-terminal system with a non-driven but possibly energy-dependent central scattering region. Time-dependent driving is applied to the right contact, modeled via Floquet coefficients in the right lead region. Contact L at which the current and noise are measured is hotter than contact R (before the driven region, indicated in blue).}
    \label{fig:example}
\end{figure}
To demonstrate the characteristics and predictiveness of the developed noise bounds, we finally present a simple example of a two-terminal single-channel system in which only one of the contacts is driven by a time-dependent voltage, see Fig.~\ref{fig:example}. 
Specifically, we choose L to be the contact where the current and noise are measured, with $T_\mathrm{L}>T_\mathrm{R}\equiv T$ and $\mu_\mathrm{L}=\bar{\mu}$, while R is subject to a time-dependent bias voltage, such that $\mu_\mathrm{R}(t)=\bar{\mu}+qV_\mathrm{R}^{\mathrm{dc}}+qV_\mathrm{R}^{\mathrm{ac}}(t)$.  The constant part of the potentials enters the Fermi functions of the two contacts. 
The scattering matrix of this system is given by 
\begin{align}
\begin{split}
        \tilde{s}_\mathrm{LR}(E,E_k) & = s_\mathrm{LR}(E)c_{-k}\\
     \tilde{s}_\mathrm{LL}(E,E_k) & = s_\mathrm{LL}(E)\delta_{k0}.
\end{split}
\end{align}
Here, the Floquet coefficients used to model the ac part of the potential in R are given by $c_{-k}$, see also Appendix~\ref{app:potentials}.
The scattering matrix $s_\mathrm{LR}(E)$ of the central region depends on one energy only, since the driving is applied to the contacts---or equivalently the leads---only.
As a result of this, the linear conductance $G$ appearing on the left-hand sides of the bounds, Eqs.~\eqref{eq:fullbound} and \eqref{eq:intersection_bound}, is the same for the driven and for the static case, see also Eq.~\eqref{eq:G_example} in Appendix~\ref{app:twoterminal_expressions}. Therefore, the excess noise on the left-hand side of the bounds indeed always represents the difference between the full nonequilibrium noise and the equilibrium noise as given by the fluctuation-dissipation theorem. We define the transmission probability of the central region as $D(E)=|s_\mathrm{LR}(E)|^2$. Here, we will choose two examples, namely a constant transmission $D(E)=D_0$ or an energy filter with a transmission of boxcar-shape.

\begin{figure}[tb]
\includegraphics[width=0.95\linewidth]{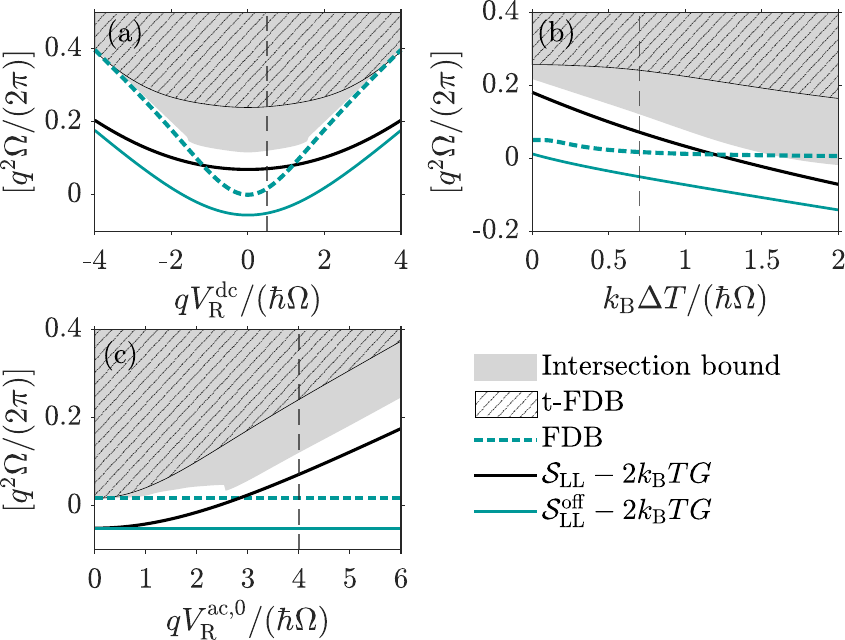}
\caption{ 
 Excess noise, $\mathcal{S}_\mathrm{LL}-2k_\mathrm{B}TG$, see Eqs.~\eqref{eq:SLL_example} and \eqref{eq:G_example}, for a time-dependently driven system with constant transmission $D_0$ compared to the analogous system where the driving is switched off. Bounds on the excess noise are marked by filled regions for the driven case and as a dotted line when the driving is switched off. We show all results (a) as a function of the stationary bias voltage $V_\mathrm{R}^\mathrm{dc}$ at $k_\mathrm{B}\Delta T=0.7\hbar \Omega$ and $qV_\mathrm{R}^{\mathrm{ac},0}=4\hbar\Omega$ (indicated by dashed vertical lines in (b) and (c)); (b) as a function of $\Delta T$ for $V_\mathrm{R}^\mathrm{dc}=0.5\hbar \Omega$ and $qV_\mathrm{R}^{\mathrm{ac},0}=4\hbar\Omega$ (indicated by dashed vertical lines in (a) and (c)); and (c) as a function of $V_\mathrm{R}^{\mathrm{ac},0}$ at $V_\mathrm{R}^\mathrm{dc}=0.5\hbar \Omega$  and $k_\mathrm{B}\Delta T=0.7\hbar \Omega$ (indicated by dashed vertical lines in (a) and (b)). In all panels, we furthermore fix $k_\mathrm{B} T=0.3\hbar\Omega$ and $D_0=0.1$. 
\label{fig:harmonic_D0}}
\end{figure}

\subsection{Harmonic driving}\label{subsec:harmonic}

The results for the noise of the most simple case, namely in the presence of a cosine-shaped harmonic driving $V_\mathrm{R}^\mathrm{ac,har}=V_\mathrm{R}^{\mathrm{ac},0}\cos(\Omega t)$ and for constant transmission $D(E)=D_0$ of the central conductor are shown in Fig.~\ref{fig:harmonic_D0}. We plot the noise as a function of the stationary biases for a fixed driving frequency and fixed driving amplitude in panels (a) and (b) and for fixed stationary biases as a function of the driving amplitude in (c). 
Black-striped regions indicate the noise values that are forbidden by the t-FDB, namely the time-dependent fluctuation-dissipation bound~\eqref{eq:fullbound}, and gray-shaded regions those forbidden by the intersection bound~\eqref{eq:intersection_bound}. 
We observe that the intersection bound is here always tighter\footnote{In other words, for \textit{constant} transmission and the driving specifically chosen in Fig.~\ref{fig:harmonic_D0}, all regions excluded by the t-FDB are also excluded by the intersection bound but not vice versa.}, in particular for small stationary biases and large driving amplitudes. This is to be expected since at low bias the hot reference distribution and the full \textit{effective}, driven distribution intersect several times and---by construction---the intersection bound selects intervals in which the two distributions can be directly compared to each other. The intersection bound furthermore shows sharp steps, which are particularly visible as functions of the driving amplitude, panel (c), and are directly connected to crossings between distribution functions (see a detailed discussion of this in Appendix~\ref{app:crossings}).  

We also show the result in the absence of time-dependent driving, here indicated by a superscript ``off", for comparison. As expected, the fluctuation-dissipation bound found in Ref.~\cite{Tesser2024May} for systems in the \textit{absence of ac-driving}, here denoted by ``FDB", breaks down when the frequency and the amplitude of the ac-driving are sufficiently large compared to the stationary biases. Instead, for large stationary biases, the ac-driving becomes less relevant as expected, and the noise as well as the bounds in the ac-driven and non-ac-driven cases approach each other.

\begin{figure}
\includegraphics[width=0.95\linewidth]{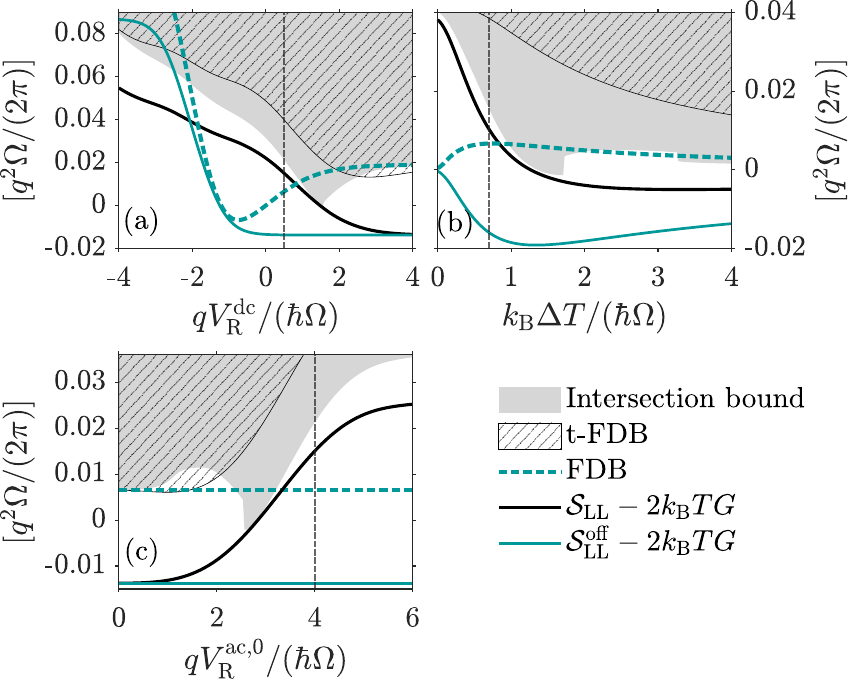}
\caption{
Excess noise, $\mathcal{S}_\mathrm{LL}-2k_\mathrm{B}TG$, see Eqs.~\eqref{eq:SLL_example} and \eqref{eq:G_example}, for a time-dependently driven system with boxcar-shaped transmission $D_\mathrm{box}(E)$, with $D_0=0.1$, $E_0=-2\hbar\Omega$ and $w=1.5\hbar\Omega$, see Eq.~\eqref{eq:boxcar}, compared to the analogous system where the driving is switched off. All other parameters are chosen as in Fig.~\ref{fig:harmonic_D0}.
\label{fig:harmonic_box}}
\end{figure}

Until here, we have focused on a situation where the central conductor uniformly transmits or reflects particles incident at any energy. We now investigate the predictiveness of the bounds when an energy-dependent transmission filters particles at energies where the time-dependent driving is particularly relevant.
In Fig.~\ref{fig:harmonic_box}, we show the fluctuations as well as their fluctuation-dissipation and intersection bounds for the case of harmonic driving using a boxcar-shaped transmission probability for the central conductor,   
\begin{equation}
    D_\mathrm{box}(E)=D_0\left(\Theta\left[E-E_0+\frac{w}{2}\right]-\Theta\left[E-E_0-\frac{w}{2}\right]\right),\label{eq:boxcar}
\end{equation}
where $E_0$ is the center of the boxcar and $w$ its width.
Due to the choice of the energy-filter position, the behavior of the noise is no longer symmetric around zero voltage bias, see Fig.~\ref{fig:harmonic_box}(a).
The energy filtering also reveals parameter regimes where the bounds are (close to) saturated, which strongly differ for the case of time-dependent driving compared to the static case (indicated by ``off").
This is illustrated in, for instance, panel (a): at a positive voltage bias of approximately $V_\mathrm{R}^\mathrm{dc}\approx 1.5\hbar\Omega$ in particular, the intersection bound for the driven case saturates, whereas at a negative bias voltage of $V_\mathrm{R}^\mathrm{dc}\approx-\hbar\Omega$ the FDB for the static case saturates. 
That energy filtering can saturate the bounds is consistent with previous findings~\cite{Tesser2024May} in the stationary case, where the bounds were shown to approach equality for weak transmission and when transport takes place in energy intervals where the hot reference distribution is close to a constant value. 
More details are provided in Appendix~\ref{app:crossings}.

Furthermore, similarly to Fig.~\ref{fig:harmonic_D0}, there are extended parameter regimes in which the excess noise of the driven system breaks the FDB for the static case. In particular, it can be achieved for negative biases [Fig.~\ref{fig:harmonic_box}(a)], at low temperature bias [Fig.~\ref{fig:harmonic_box}(b)], or also at large driving amplitude [Fig.~\ref{fig:harmonic_box}(c)].
This clearly shows the need for a dedicated bound for the time-dependent driving. Additionally, panel (a) shows that at large negative voltage biases the static excess noise exceeds both bounds for the time-dependent case.

In contrast with Fig.~\ref{fig:harmonic_D0}, the choice of an energy-dependent transmission function in Fig.~\ref{fig:harmonic_box} demonstrates the absence of a hierarchy, not only between the two different bounds for the time-dependently driven case, but also between the bound for the static case and the bounds for the time-dependently driven case.
Indeed, while we have seen that the intersection bound is typically tighter, in panels (a) and (c) the t-FDB is lower than the intersection bound for some parameter regimes, namely for large voltage bias and driving amplitude, respectively.

\subsection{Comparison of different ac-driving shapes}

\begin{figure}[t]
    \centering
    \includegraphics[width=0.9\linewidth]{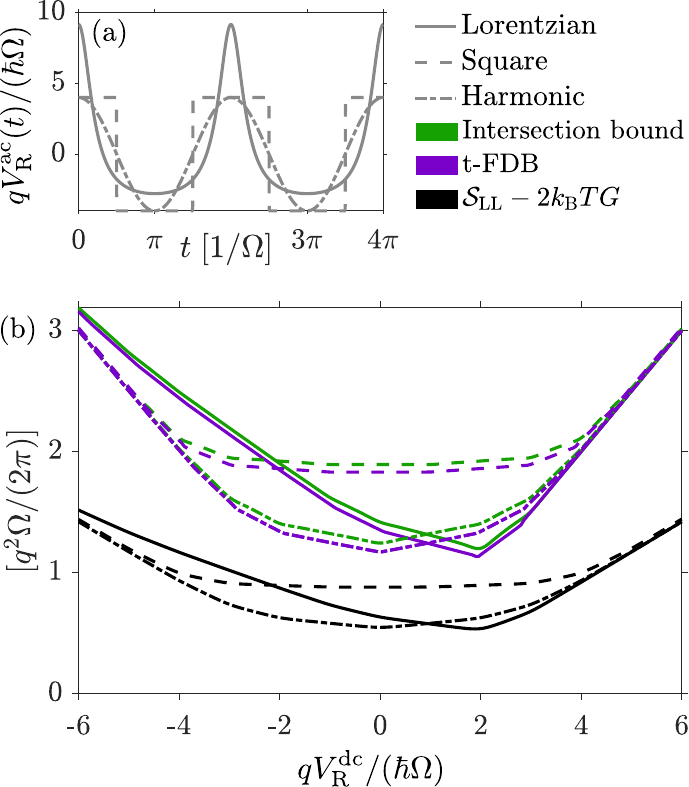}
    \caption{Comparison of the effects of different driving schemes, as indicated in (a). Panel (b) shows the excess noise and noise bounds as function of the stationary bias for these different driving shapes. We choose the parameters  $k_\mathrm{B}T=0.01\hbar\Omega$, $k_\mathrm{B}\Delta T=0.1\hbar\Omega$, $D\equiv D_0=0.5$, and  $q V_\mathrm{R}^{\mathrm{ac},0}=4\hbar\Omega$.  We furthermore have $\sigma=0.1 \mathcal{T}$ for the width of the Lorentzian pulses, see Appendix~\ref{app:potentials} for details about the signal shapes.}
    \label{fig:diff_driving}
\end{figure}

As a next step, we investigate how different driving shapes impact the noise and the bounds. 
We choose three specific driving potentials $V_\mathrm{R}^{\mathrm{ac}}(t)$ as examples to demonstrate the implications of the discovered bounds. They are (i) the most simple harmonic drive, with cosine shape $qV_\mathrm{R}^\mathrm{ac,har}(t)$ as discussed in the previous Sec.~\ref{subsec:harmonic}, (ii) a Lorentzian drive, $qV_\mathrm{R}^\mathrm{ac,Lor}(t)$, which in the zero-temperature limit is known to have minimal excess noise~\cite{Ivanov1997Sep,Keeling2006Sep,Dubois2013Oct}, and (iii) a square drive, $qV_\mathrm{R}^\mathrm{ac,squ}(t)$, which instead has been shown to result in large excess noise~\cite{Bertin-Johannet2022Mar}. These three signals are shown in Fig.~\ref{fig:diff_driving}(a). Details about the specific shape of the potentials and the Floquet coefficients modeling them are provided in Appendix~\ref{app:potentials}.
We choose a low-temperature regime to highlight the characteristic features of the different drivings. Indeed, as expected, there are regions where the Lorentzian driving outperforms the harmonic driving by displaying the lowest noise level in excess to the thermal noise---here occurring at $V_\mathrm{R}^\mathrm{dc}\approx2\hbar\Omega$. Also as expected, the square drive is the driving scheme that generates the highest level of excess noise~\cite{Bertin-Johannet2022Mar}.
These properties of the noise are also reflected in the shape of the fluctuation-dissipation bound and the intersection bound. However, the bounds are far from being tight in the parameter regime chosen here to highlight the differences in the excess noise between different driving scheme. 
This large difference between excess noises and their bounds can be ascribed to the fact that we here choose an energy-independent transmission of significant magnitude (where we expect the t-FDB to be least tight, as confirmed by the analysis of the static case~\cite{Tesser2024May}), in order not to mix features of driving and of energy filtering and to evidence the difference in the driving schemes.

The characteristics associated to the different driving schemes disappear rather rapidly with increasing voltage bias or with increasing temperature bias (not shown here). For large biases the noises and bounds would also approach the results for the non-driven case, see Fig.~\ref{fig:harmonic_D0}(a) for comparison.  This demonstrates that---in the simple two-terminal setup with ac-voltage bias driving---the shape of the driving signal has a less important impact on the noise bounds than for example the energy filtering of the central conductor discussed above in Sec.~\ref{subsec:harmonic}.

\subsection{Relation of noise bounds to power production}

Finally, we demonstrate the relation between the \mbox{t-FDB} and the dissipated powers due to static bias voltage and driving, see~\eqref{eq:largebiasbound_power}, for the example of the two-terminal conductor driven by a harmonic ac potential in the terminals, as shown in Fig.~\ref{fig:example}, and discussed in Sec.~\ref{subsec:harmonic}. The explicit expressions are given in Eq.~\eqref{eq:SLL_example} for the noise and in Eqs.~\eqref{eq:Ppot_example} and~\eqref{eq:Pdriv_example} for the power.

\begin{figure}
    \centering
    \includegraphics[width=0.95\linewidth]{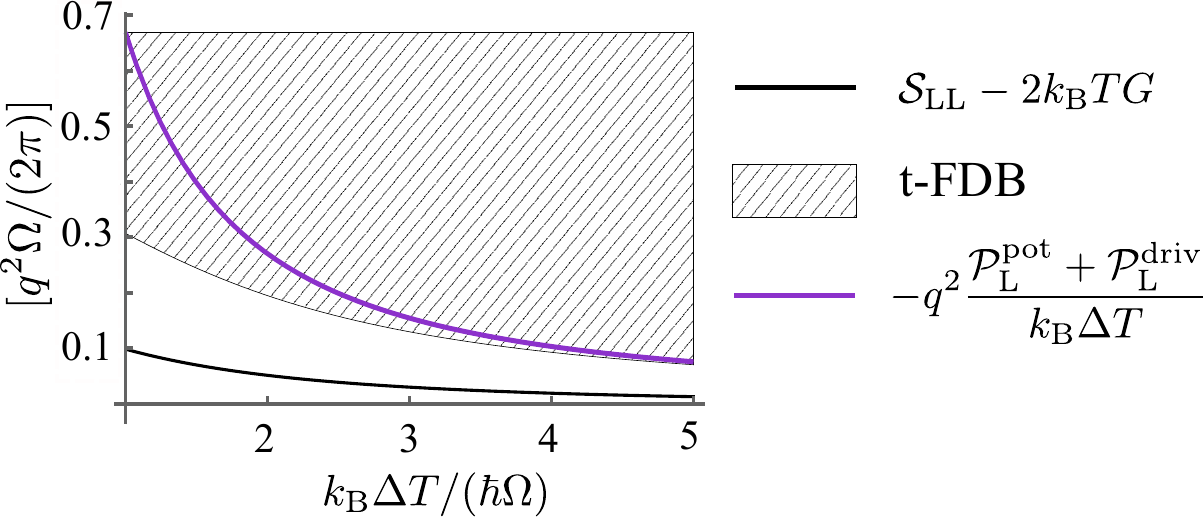}
    \caption{Excess noise, t-FDB [see Eq.~\eqref{eq:fullbound}], and t-FDB in the limit of large temperature bias, $\Delta T\gg\Delta\mu+\hbar\Omega k$ expressed in terms of dissipated power [Eq.~\eqref{eq:largebiasbound_power}]. The driving has cosine shape with $qV_\mathrm{R}^{\mathrm{ac},0}=2\hbar\Omega$. The boxcar-shaped transmission probability has $D_0=1$, $E_0=3\hbar\Omega$ and $w=\hbar\Omega$. We furthermore have $qV_\mathrm{R}^\mathrm{dc}=3\hbar\Omega$ and $k_\mathrm{B}T=\hbar\Omega$.
    }
    \label{fig:power}
\end{figure}

The excess noise of the driven system is shown in Fig.~\ref{fig:power} together with the full t-FDB~\eqref{eq:fullbound} and its limit for large temperature differences which directly relates to the dissipated powers~\eqref{eq:largebiasbound_power}. We see that the large $\Delta T$-approximation, namely the sum of dissipated powers divided by the temperature difference, approaches the t-FDB already at temperature differences of the order of $\Delta T\approx 3.5\hbar\Omega$. While the excess noise is suppressed, tending to zero, in this limit, the full t-FDB as well as its approximation in terms of dissipated powers still describes the overall behaviour and yields a reasonable estimate of the excess noise. This shows that the approximated t-FDB bound, which does not require any information about the Floquet decomposition of the current, constitutes a good estimate for the noise in large parameter regimes.

\section{Conclusions}

We have derived bounds on the excess noise of generic time-dependently driven electronic coherent conductors, which are furthermore subject to possibly large static voltage and temperature biases. The first of these bounds, referred to as the fluctuation-dissipation bound for time-dependently driven systems~\eqref{eq:fullbound}, admits an intuitive interpretation in terms of power dissipated by the time-dependent driving and the static biases~\eqref{eq:largebiasbound_power}, which is particularly predictive in the regime of large temperature bias. A second bound, referred to as intersection bound~\eqref{eq:intersection_bound}, relies on knowledge about the crossing points between (effective) distribution functions. 
We foresee that this intersection bound is of use for the characterization of noise in systems with generic nonthermal distributions, i.e. not necessarily stemming from time-dependent driving~\cite{Danielsson2025Aug}.
In the absence of driving (or other nonthermal effects), both bounds tend to the static fluctuation-dissipation bound, previously developed in Ref.~\cite{Tesser2024May}.

We have demonstrated the validity of the bounds and highlighted their characteristic features for the simple, but experimentally relevant~\cite{Gabelli2008Jan,Dubois2013Oct,Glattli2017Mar}, example case of a two-terminal out-of equilibrium conductor subject to an ac bias voltage.
However, the validity of the presented bounds extends far beyond this case. We expect the bounds to be useful to understand and to constrain, e.g., the noise of periodically operated heat engines~\cite{Kosloff2014Apr} and of single-electron sources in the presence of (accidental) temperature gradients.

\acknowledgments
We thank Rafael Sánchez and Elsa Danielsson for helpful discussions throughout the duration of the project, and Matteo Acciai for providing useful feedback on the manuscript. We gratefully acknowledge funding from the Knut and Alice Wallenberg Foundation via the Fellowship program and from the European Research Council (ERC) under the European Union’s Horizon Europe research and innovation program (101088169/NanoRecycle) (L.T. and J.S.) as well as from the Spanish Ministerio de Ciencia e Innovaci\'on via grants No. PID2019-110125GB-I00, No. PID2022-142911NB-I00 and No. PID2024-157821NB-I00 (J.B.).

\appendix

\section{Properties of Floquet scattering matrix}\label{app:properties}
In this appendix, we show some of the important properties of the scattering matrix and the Floquet coefficients.

\subsection{Floquet coefficients}\label{app:c_alpha}

The Floquet coefficients fulfill some important properties. To show them, we start from the definitions in Eqs.~\eqref{eq:Floquet_coefficients}.
First, we demonstrate the sum rule
\begin{eqnarray}
  &&  \sum_k c_{\alpha(k+p)}^*c_{\alpha(k+\ell)} = \nonumber\\
  & = & \int_0^\mathcal{T}\!\!\frac{dt}{\mathcal{T}}\int_0^\mathcal{T}\!\!\frac{dt'}{\mathcal{T}}\sum_{k}e^{i\phi_\alpha(t')}e^{-i(k+p)\Omega t'}e^{-i\phi_\alpha(t)}e^{i(k+\ell)\Omega t}\nonumber\\
& = &   \int_0^\mathcal{T}\frac{dt}{\mathcal{T}}e^{-i(p-\ell)\Omega t}  = \delta_{p\ell}\label{eq:Floquet1}
\end{eqnarray}
Furthermore, we demonstrate a relation needed to express the components of the driving potential in terms of Floquet coefficients
\begin{eqnarray}
  &&  \sum_k k c_{\alpha(k-p)}^*c_{\alpha k}\nonumber\\ & = & \int_0^\mathcal{T}\frac{dt'}{\mathcal{T}}\int_0^\mathcal{T}\frac{dt}{\mathcal{T}}e^{i\phi_\alpha(t')-i\phi_\alpha(t)}\sum_{k}ke^{ik\Omega t}e^{-i(k-p)\Omega t'}.
    \nonumber
\end{eqnarray}
Rewriting the factor $k$ as a derivative of the exponential $e^{ik\Omega t}$ and integrating by parts while using the periodicity of the driving potential, we find 
\begin{eqnarray}
     \sum_k k c_{\alpha(k-p)}^*c_{\alpha k} 
     & = & \int_0^\mathcal{T}\frac{dt'}{\mathcal{T}}\int_0^\mathcal{T}\frac{dt}{\mathcal{T}}e^{i\phi_\alpha(t')-i\phi_\alpha(t)}\nonumber\\
     &&\times \sum_{k}\frac{q}{\hbar\Omega}V^\mathrm{ac}_\alpha(t)e^{ik\Omega t}e^{-i(k-p)\Omega t'}
    \nonumber\\ 
    & = & \frac{q}{\hbar\Omega}\int_0^\mathcal{T}\frac{dt}{\mathcal{T}}\sum_nV_{\alpha  n}e^{-in\Omega t}e^{ip\Omega t}\nonumber\\
    & = & \frac{q}{\hbar\Omega}V_{\alpha p}\label{eq:V_floquet}\ .
\end{eqnarray}
This result directly implies
\begin{eqnarray}
    \sum_k k |c_{\alpha, k}|^2 & = & 0 \ .
\end{eqnarray}

\subsection{Unitarity of Floquet scattering matrix}\label{app:unitarity}

The Floquet scattering matrix fulfills the unitarity condition
\begin{equation}
    \sum_{\beta,p}\tilde{t}^\dagger_{\alpha\beta}(E_\ell,E_p)\tilde{t}_{\alpha\beta}(E_k,E_p)=\delta_{k\ell}\mathbf{1}_\alpha
\end{equation}
here written for the submatrices $\tilde{t}_{\alpha\beta}$ with $[\tilde{t}_{\alpha\beta}]_{nm}=\tilde{s}_{\alpha n,\beta m}$, with $\mathbf{1}_\alpha$ the unit matrix of dimension $N_\alpha$.

\section{Power due to driving and static biases}\label{app:power}

 We need the expressions for the power in order to interpret the t-FDB, namely the fluctuation-dissipation bound with time-dependent driving. We start by writing down the energy current in the driven system \textit{flowing into contact $\alpha$}, which is given by
\begin{eqnarray}
 I_\alpha^E & = &  \int dE\ \frac{E}{h}\sum_{\beta,k}\Tr{}{\tilde{t}^\dagger_{\alpha\beta}(E,E_{k})\tilde{t}_{\alpha\beta}(E,E_{k})} \nonumber\\
 &&\left(f_\beta(E_{k})- f_\alpha(E)\right)\ .\label{eq:Ecurrent}
\end{eqnarray}
The sum over all energy-currents yields the total power provided by the driving and dissipated in the system. This is equivalent to \textit{minus} the power received by the driving. To find this received power, $\mathcal{P}^\mathrm{driv}$, we first rewrite the negative of the sum over energy currents using the unitarity of the scattering matrix
\begin{eqnarray}
 &&  \mathcal{P}^\mathrm{driv}= -\sum_\alpha I^E_\alpha = -\sum_{\alpha}\int dE\ \frac{E}{h}\times\nonumber\\
 && \times\left[\sum_{\beta,k}\Tr{}{\tilde{t}^\dagger_{\alpha\beta}(E,E_k)\tilde{t}_{\alpha\beta}(E,E_{k})}f_\beta(E_{k})- N_\alpha f_\alpha(E)\right]\nonumber\\
   & & = \sum_{\alpha,\beta,k}\int dE\ \frac{E}{h}\left(\Tr{}{\tilde{t}^\dagger_{\beta\alpha}(E_k,E)\tilde{t}_{\beta\alpha}(E_{k},E)}f_\alpha(E) \right.\nonumber\\
   && - \left.\Tr{}{\tilde{t}^\dagger_{\alpha\beta}(E,E_k)\tilde{t}_{\alpha\beta}(E,E_{k})}f_\beta(E_{k})\right)\label{eq:Ecurrent_step1}\ .
\end{eqnarray}
As a next step we shift the energy $E\rightarrow E_{-k}$, and swap the indices $k\rightarrow -k$ and $\alpha\leftrightarrow\beta$, which results in
\begin{eqnarray}
 && \sum_{\alpha,\beta,k}\int dE\ \frac{E_k}{h}\left(\Tr{}{\tilde{t}^\dagger_{\alpha\beta}(E,E_k)\tilde{t}_{\alpha\beta}(E,E_k)}f_\beta(E_k)\right.\nonumber\\
 &&\left. -\Tr{}{\tilde{t}^\dagger_{\beta\alpha}(E_k,E)\tilde{t}_{\beta\alpha}(E_k,E)}f_\alpha(E)\right)\label{eq:Ecurrent_step2}\ .
\end{eqnarray}
Summing now the identical expressions \eqref{eq:Ecurrent_step1} and \eqref{eq:Ecurrent_step2} and dividing by 2, we find the power $\mathcal{P}^\text{driv}$ as 
\begin{eqnarray}
 &&  \mathcal{P}^\text{driv}  = \nonumber\\
 && \sum_{\alpha,\beta,k}\int dE\ \frac{k\Omega}{4\pi}\left(\Tr{}{\tilde{t}^\dagger_{\alpha\beta}(E,E_k)\tilde{t}_{\alpha\beta}(E,E_k)}f_\beta(E_k)\nonumber\right.\\
 && \left.-\Tr{}{\tilde{t}^\dagger_{\beta\alpha}(E_k,E)\tilde{t}_{\beta\alpha}(E_k,E)}f_\alpha(E)\right)\ .\label{eq:powercomponents}
\end{eqnarray}
In the same way, one can derive the power received by the time-reversed protocol, exploiting the behavior of the Floquet scattering matrix under time-reversal, $\left[\tilde{t}_{\alpha\beta}(E,E_k)\right]^\mathrm{tr}=\tilde{t}_{\beta\alpha}(E_k,E)$, see~\cite{Moskalets2005Jul} and page 73 in Ref.~\cite{Moskalets2011Sep}. One then finds
\begin{eqnarray}
  && \mathcal{P}^\text{driv,tr} := \nonumber\\ 
  &&\sum_{\alpha,\beta,k} \int dE\ \frac{k\Omega}{4\pi}\left(\Tr{}{\tilde{t}^\dagger_{\beta\alpha}(E_k,E)\tilde{t}_{\beta\alpha}(E_k,E)}f_\beta(E_k)+\right.\nonumber\\
  &&\left. -\Tr{}{\tilde{t}^\dagger_{\alpha\beta}(E,E_k)\tilde{t}_{\alpha\beta}(E,E_k)}f_\alpha(E)\right)\ .
\end{eqnarray}
We can proceed in a similar way to find the power generated in contact $\alpha$ due to a current flowing in the presence of static voltage biases. Thus, we start from a sum over all heat currents in contact $\alpha$, which are given by $J_\alpha=I_\alpha^E-\mu_\alpha I_\alpha^N$. We have already shown how the sum over all energy currents yields the power dissipated due to the driving or received by the driving fields. The power \textit{produced} by the part of the energy current stemming from the chemical work is given by
\begin{eqnarray}
    \mathcal{P}^\text{pot}& = & \sum_\alpha\mu_\alpha I_\alpha^N = - \sum_{\alpha,\beta,k}\frac{\mu_\alpha}{h}\int dE\ \\
    && \Tr{}{\tilde{t}^\dagger_{\alpha\beta}(E,E_k)\tilde{t}_{\alpha\beta}(E,E_k)}\left(f_\beta(E_k)-f_\alpha(E)\right)\ .\nonumber
\end{eqnarray}
 For the power generated in the time-reversed situation, we instead find
\begin{eqnarray}
        \mathcal{P}^\text{pot,tr} & = &  \sum_{\alpha,\beta,k}\frac{\mu_\alpha}{h}\int dE\ \Tr{}{\tilde{t}^\dagger_{\beta\alpha}(E_k,E)\tilde{t}_{\beta\alpha}(E_k,E)}\nonumber\\
        && \left(f_\beta(E_k)-f_\alpha(E)\right)\nonumber\\
        & = &  \sum_{\alpha,\beta,k}\frac{\mu_\beta}{h}\int dE\ \Tr{}{\tilde{t}^\dagger_{\alpha\beta}(E,E_k)\tilde{t}_{\alpha\beta}(E,E_k)}\nonumber\\
        &&\left(f_\alpha(E)-f_\beta(E_k)\right)\label{eq:Ppot_rev}
\end{eqnarray}
where we swapped indices and shifted energies in the second line of Eq.~\eqref{eq:Ppot_rev}.

\section{Derivation of the t-FDB}\label{app:inequalities}
To derive the t-FDB, namely the fluctuation-dissipation bound in the presence of time-dependent driving, we have split the noise in the three contributions $\mathcal{S}_{\alpha\alpha}^{(0)}$, $\mathcal{S}_{\alpha\alpha}^{(2)}$, and $\mathcal{S}_{\alpha\alpha}^{(4)}$ in Eq.~\eqref{eq:S_contributions}.
We have written $\mathcal{S}_{\alpha\alpha}^{(2)}$ in terms of the linear conductance and a contribution that partially cancels out with $\mathcal{S}_{\alpha\alpha}^{(0)}$ in Eq.~\eqref{eq:S23_contribution}. 
Here, we will show how to rewrite and estimate the remaining contributions from $\mathcal{S}_{\alpha\alpha}^{(0)}$ and in particular from $\mathcal{S}_{\alpha\alpha}^{(4)}$. 
First, we split the contribution $\mathcal{S}_{\alpha\alpha}^{(4)}$ into two pieces, one linear in the Fermi functions $\mathcal{S}_{\alpha\alpha}^{(4,1)}$ and one quadratic in the Fermi functions $\mathcal{S}_{\alpha\alpha}^{(4,2)}$.  
The scalar product nature of the latter,
\begin{eqnarray}
    &&  \mathcal{S}^{(4,2)}_{\alpha\alpha} =   -\frac{q^2}{h} \int dE  
    f_\beta(E_{k})f_\gamma(E_{p})\times \label{eq:CS1}  \\ 
    &&\Tr{}{\tilde{t}_{\alpha\beta}(E_{\ell},E_{k})\tilde{t}^\dagger_{\alpha\beta}
    (E,E_{k})\tilde{t}_{\alpha\gamma}(E,E_{p})\tilde{t}^\dagger_{\alpha\gamma}(E_{\ell},E_{p})},\nonumber
\end{eqnarray}
allows us to use the Cauchy-Schwarz inequality, such that 
\begin{eqnarray}
 &&\mathcal{S}^{(4,2)}_{\alpha\alpha}\\
 &&\leq
-\frac{q^2}{h N_\alpha} \int dE  \left| \Tr{}{\tilde{t}_{\alpha\beta}(E,E_{k})\tilde{t}^\dagger_{\alpha\beta}
    (E,E_{k})}f_\beta(E_{k})\right|^2\nonumber\ .
\end{eqnarray}
To be able to further treat this term, we add and subtract $f_\alpha(E)$ from $f_\beta(E_{k})$ to find
\begin{eqnarray}
 &&\mathcal{S}^{(4,2)}_{\alpha\alpha}\leq -\frac{q^2}{h} \int dE  N_\alpha f_\alpha(E)f_\alpha(E)\nonumber\\
&&-\frac{q^2}{hN_\alpha} \int dE    \left| \mathrm{Tr}\left\{\tilde{t}_{\alpha\beta}(E,E_{k})\tilde{t}^\dagger_{\alpha\beta}
    (E,E_{k})\right\}\right.\times\nonumber\\
    && \left.\hspace{2cm}\times\left(f_\beta(E_{k})-f_\alpha(E)\right)\right|^2\nonumber\\
&&-2\frac{q^2}{h} \int dE  \Tr{}{\tilde{t}_{\alpha\beta}(E,E_{k})\tilde{t}^\dagger_{\alpha\beta}
    (E,E_{k})}\times\nonumber\\
    && \hspace{2cm}\times f_\alpha(E)\left(f_\beta(E_{k})-f_\alpha(E)\right)\ .  \label{eq:S5plus}
\end{eqnarray}
We note here that the first term will cancel with parts of $\mathcal{S}^{(0)}$ respectively $\mathcal{S}^{(2)}$, while the second one is finite but always negative. As a next step, before summing together all contributions, we now analyze $\mathcal{S}^{(4,1)}$, which reads
\begin{eqnarray}
    &&  \mathcal{S}^{(4,1)}_{\alpha\alpha} =   \frac{q^2}{h} \int dE  
    f_\beta(E_{k})\times \label{eq:CS1}  \\ 
    &&\Tr{}{\tilde{t}_{\alpha\beta}(E_{\ell},E_{k})\tilde{t}^\dagger_{\alpha\beta}
    (E,E_{k})\tilde{t}_{\alpha\gamma}(E,E_{p})\tilde{t}^\dagger_{\alpha\gamma}(E_{\ell},E_{p})}\nonumber\\
    &&= \frac{q^2}{h} \int dE  
    f_\beta(E_{k})\Tr{}{\tilde{t}_{\alpha\beta}(E,E_{k})\tilde{t}^\dagger_{\alpha\beta}
    (E,E_{k})}\nonumber\ .
\end{eqnarray}
Here we again subtract and add $f_\alpha(E)$ to find
 \begin{eqnarray}
\mathcal{S}^{(4,1)}_{\alpha\alpha} & = & \frac{q^2}{h} \int dE  
    N_\alpha f_\alpha(E)\\
    && + \frac{q^2}{h} \int dE  \left(f_\beta(E_{k})- f_\alpha(E)\right)\times\nonumber\\
 &&\times   \Tr{}{\tilde{t}_{\alpha\beta}(E,E_{k})\tilde{t}^\dagger_{\alpha\beta}
    (E,E_{k})}\nonumber\ .
\end{eqnarray}
Now summing $\mathcal{S}^{(0)}_{\alpha\alpha}$, $\mathcal{S}^{(2)}_{\alpha\alpha}$, and $\mathcal{S}^{(4,1)}_{\alpha\alpha}$ to the inequality developed starting from $\mathcal{S}^{(4,2)}_{\alpha\alpha}$ in~\eqref{eq:S5plus}, we reach the bound presented in~\eqref{eq:start_bound} in the main text.

\section{Alternative intersection bound}\label{app:intersection}
In Sec.~\ref{subsec:intersection}, we have presented a noise bound that is based on intersections between \textit{effective} distribution functions. The shape of this bound hence heavily depends on how the effective distribution functions are defined. An alternative way to write a bound, in which the current-like shape of the bound contributions is highlighted, is by introducing the effective distribution function
\begin{equation}
    f^\star_\alpha(E) = \frac{\sum\limits_{(\beta, k)\notin \{(\alpha,0)\}}\Tr{}{\tilde{t}_{\alpha\beta}
    (E,E_{k})\tilde{t}^\dagger_{\alpha\beta}(E,E_{k})}f_\beta(E_{k})}{\sum\limits_{(\beta, k)\notin \{(\alpha,0)\}}\Tr{}{\tilde{t}_{\alpha\beta}
    (E,E_{k})\tilde{t}^\dagger_{\alpha\beta}(E,E_{k})}}\ ,
\end{equation}
which is well-defined as long as $\Tr{}{\tilde{t}_{\alpha\beta}(E,E_{k})\tilde{t}^\dagger_{\alpha\beta}(E,E_{k})}\neq 0$ for at least one $(\beta, k)\notin \{(\alpha,0)\}$.
This modified distribution function $f^\star_\alpha(E)$, when defined on the entire energy interval, has $M$ crossings with the hottest distribution $f_\alpha(E)$, which are the same crossings as the one between $f_\alpha^\bigstar(E)$ and $f_\alpha(E)$. With this, we can write the intersection bound as 
\begin{eqnarray}
&&   \mathcal{S}_{\alpha\alpha} -  2\kB T_\alpha\sum_{\beta\neq\alpha}G_{\alpha\beta} \leq q^2 \sum_{n=0}^{\frac{M-1}{2}}(1-2f_\alpha(\varepsilon_{2n+1}))\nonumber\\
&&\times\int_{\varepsilon_{2n}}^{\varepsilon_{2n+2}} \frac{dE}{h} D_{\alpha}(E) \left(f^\star_\alpha(E)-f_\alpha(E)\right) . \label{eq:2nd_intersection_bound}
\end{eqnarray}
Here, the factor $D_\alpha(E)=N_\alpha-\Tr{}{t^\dagger_{\alpha\alpha}(E)t_{\alpha\alpha}(E)}$ can be understood as a transmission probability for a current resulting from the difference in occupations $f^\star_\alpha(E)-f_\alpha(E)$ in the interval $[\varepsilon_{2n},\varepsilon_{2n+2}]$.

\section{Explicit expressions for the two-terminal ac-driven conductor}\label{app:twoterminal_expressions}

In the simple two-terminal system subject to an ac voltage bias, as treated in Sec.~\ref{sec:plots}, the full noise is found to be~\cite{Battista2014Aug}
\begin{align}
    &\mathcal{S}_{\mathrm{LL}}  =  \frac{q^2}{h}\int dE \Big[D(E)^2f_\mathrm{L}(E)f^-_\mathrm{L}(E)\label{eq:SLL_example}\\
 & +D(E)D(E_\ell)c_{l-k}c^*_{-k}c_{l-p}c^*_{-p}f_R(E_k)f_\mathrm{R}^-(E_p)\nonumber\\
 & + D(E)(1-D(E))\left(f_\mathrm{L}(E)\tilde{f}_\mathrm{R}^-(E_k)+f_\mathrm{L}^-(E)\tilde{f}_\mathrm{R}(E_k)\right)\Big]\nonumber
\end{align}
where the first two lines are the interference contributions to the noise, which are equal to the thermal noise in the absence of driving and the last line is the transport contribution to the noise. Here, we have defined 
\begin{equation}\label{eq:ftilde}
\tilde{f}_\mathrm{R}(E) \definition \sum_{k}|c_{-k}|^2f_\mathrm{R}(E_k).
\end{equation}
To establish the fluctuation-dissipation and intersection bounds, we write the conductance~\eqref{eq:lin_conductance}, 
\begin{eqnarray}
    G \equiv G_\mathrm{LR} &=& \frac{q^2}{h\kB T_\mathrm{L}}\int dE\ D(E)f_\mathrm{L}(E)f^-_\mathrm{L}(E),\label{eq:G_example}
\end{eqnarray}
which is independent of the driving in the leads and simply proportional to the thermal noise of contact L.
The current components of Eq.~\eqref{eq:current_components} are given by
\begin{equation}
    I_{\mathrm{LR,k}}=q\int \frac{dE}{h}D(E)|c_{-k}|^2\left(f_\mathrm{R}(E_{k})-f_\mathrm{L}(E)\right).
\end{equation}
Starting from this expression, we calculate the dissipated powers, which enter the t-FDB~\eqref{eq:fullbound} in the limit of large temperature bias,
\begin{eqnarray}
  && 2\mathcal{P}_\mathrm{L}^\mathrm{pot}  = \label{eq:Ppot_example} \\
  && -qV_\mathrm{R}^\mathrm{dc}\int \frac{dE}{h}D(E)|c_{-k}|^2\left(f_\mathrm{R}(E_{k})-f_\mathrm{L}(E)\right) \nonumber\\
&&   2\mathcal{P}_\mathrm{L}^\mathrm{driv} = \label{eq:Pdriv_example}\\
  &&   \int \frac{dE}{h}D(E) \hbar\Omega k|c_{-k}|^2\left(f_\mathrm{R}(E_{k})-f_\mathrm{L}(E)\right) \nonumber\ .
\end{eqnarray} 
The first contribution, $\mathcal{P}^\mathrm{pot}_\mathrm{L}$, is the dissipated power due to the dc current flowing in the presence of a potential bias. In the limit of no driving or in the limit of constant transmission $D(E)\equiv D_0$, where the ac driving averages out, this contribution equals the power dissipated in a steady-state system~\cite{Tesser2024May}.
The second contribution, $\mathcal{P}^\mathrm{driv}_\mathrm{L}$, is the power that is dissipated due to the dc drive. 

Also the intersection bounds of Sec.~\ref{subsec:intersection} and Appendix~\ref{app:intersection} take a simple form for the two-terminal system of Fig.~\ref{fig:example}. 
In particular, one finds for the integral contribution of the intersection bound~\eqref{eq:intersection_bound},
\begin{align}
    &\int_{\varepsilon_{2n}}^{\varepsilon_{2n+2}} \frac{dE}{h} \left(f^\bigstar_\alpha(E)-f_\alpha(E)\right) \nonumber\\
    &\rightarrow \int_{\varepsilon_{2n}}^{\varepsilon_{2n+2}} \frac{dE}{h} D(E) \left(\tilde{f}_\mathrm{R}(E)-f_\mathrm{L}(E)\right) 
\end{align}
since $f^\bigstar_\mathrm{L}(E)$ and $\tilde{f}_\mathrm{R}(E)$ have the same crossing points with $f_\mathrm{L}(E)$ in the relevant energy intervals whenever $D(E)\neq 0$. The definition of $\tilde{f}_\mathrm{R}(E)$ is given in Eq.~\eqref{eq:ftilde}. For the example treated here, $\tilde{f}_\mathrm{R}(E)$ also equals the analytical continuation of $f^\star_\mathrm{L}(E)$ from the intersection-bound version of~\eqref{eq:2nd_intersection_bound}. In the analysis of the noise bounds based on crossings between distributions in Appendix~\ref{app:crossings}, we therefore consider crossings between $\tilde{f}_\mathrm{R}(E)$ and $f_\mathrm{L}(E)$.

\section{Driving potentials and Floquet coefficients}\label{app:potentials}

Here we provide explicit expressions for the different types of driving potentials as well as the Floquet coefficients that we used to model the pure ac-part of these potentials. 

For the case (i), where the ac driving potential in the right contact is given by a harmonic, cosine-shaped drive
\begin{subequations}
    \begin{equation}
    V_\mathrm{R}^{\mathrm{ac,har}}(t)=V_\mathrm{R}^{\mathrm{ac},0}\cos(\Omega t),
\end{equation}
the Floquet coefficients incorporating the effect of this ac component of the potential are given by Bessel functions,
\begin{equation}\label{eq:c_cos}
    c_{-k}^\mathrm{har} = J_{-k}\left(\frac{q V_\mathrm{R}^{\mathrm{ac},0}}{\hbar\Omega}\right)\ .
\end{equation}
\end{subequations}
For a Lorentzian-shaped drive with one Lorentzian-shaped voltage peak per period, case (ii), where the pure ac-component of the drive reads
\begin{subequations}
\begin{align}
V_\mathrm{R}^{\mathrm{ac,Lor}}(t)=V_\mathrm{R}^{\mathrm{ac},0}\sum_j\frac{\mathcal{T}\sigma/\pi}{(t-t_\mathrm{Lor}-j\mathcal{T})^2+\sigma^2}-V_\mathrm{R}^{\mathrm{ac},0}\ ,    
\end{align}
we instead have for its representation in terms of Floquet coefficients~\cite{Moskalets2011Sep,Dubois2013Oct,Bertin-Johannet2022Mar}
\begin{align}\label{eq:c_Lor}
c_{-k}^\mathrm{Lor} = & \int_0^1 du\left(\frac{\sin[\pi(u+i\sigma/\mathcal{T})]}{\sin[\pi(u-i\sigma/\mathcal{T})}\right)^{qV_\mathrm{R}^{\mathrm{ac},0}/\hbar\Omega}\nonumber\\
& \times\exp\left[2\pi i u\left(-k-\frac{qV_\mathrm{R}^{\mathrm{ac},0}}{\hbar\Omega}\right)\right]
\end{align}
\end{subequations}
Finally, for case (iii), the Floquet coefficients for the ac-component of a square drive with
\begin{subequations}
\begin{equation}
      V_\mathrm{R}^{\mathrm{ac,squ}}(t)=V_\mathrm{R}^{\mathrm{ac},0}\mathrm{sgn}[\cos(\Omega t)]
\end{equation} 
are given by~\cite{Bertin-Johannet2022Mar}
\begin{equation}\label{eq:c_Lor}
    c_{-k}^\mathrm{squ} = \frac{2}{\pi}\frac{qV_\mathrm{R}^{\mathrm{ac},0}/\Omega}{k^2-\left(qV_\mathrm{R}^{\mathrm{ac},0}/\Omega\right)^2}\sin\left[\frac{\pi}{2}\left(-k-qV_\mathrm{R}^{\mathrm{ac},0}/\Omega\right)\right]\ .
\end{equation}
\end{subequations}

\section{Intersections between distribution functions}\label{app:crossings}

In this appendix, we demonstrate how a number of features in the bounds, observed in Figs.~\ref{fig:harmonic_D0} and \ref{fig:harmonic_box}, can be explained by examining the relevant crossings between effective driven distributions and the reference distribution of the hot reservoir. For simplicity, we here always show  $\tilde{f}_\mathrm{R}(E)$, see Eq.~\eqref{eq:ftilde}, instead of $f^\bigstar_\mathrm{L}(E)$, see Eq.~\eqref{eq:fbigstar}. This simplifies the plots and the discussion, while the crossing points remain the same.

\subsection{Sharp features in intersection bound}

In Figs.~\ref{fig:harmonic_D0} and \ref{fig:harmonic_box}, the intersection bound displays discontinuities as function of different externally tunable parameters. This can be explained by examining how crossings between the hot reference distribution and the effective distribution of the driven contact appear and disappear as function of those parameters. 

\begin{figure}[t]
    \centering
    \includegraphics[width=0.9\linewidth]{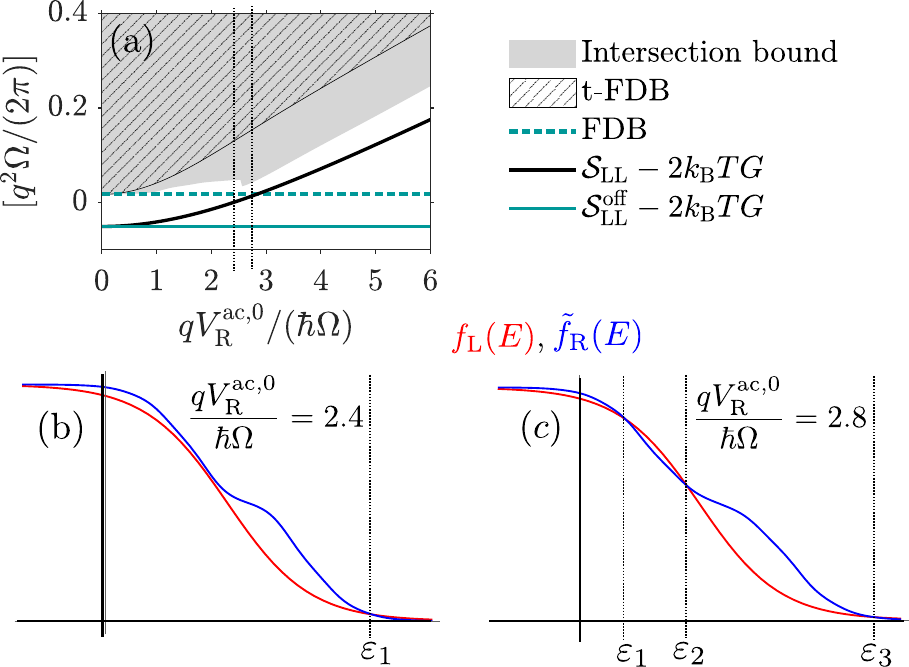}
    \caption{Excess noise and bounds for a harmonically driven two-terminal conductor with constant transmission probability [all parameters in panel (a) as in Fig.~\ref{fig:harmonic_D0}(c) in the main text]. Panels (b) and (c) show the distribution function for the hot contact (red) and the modified distribution from the cold contact due to driving (blue) at the parameter values indicated by vertical dashed lines in (a), namely at $qV_\mathrm{R}^{\mathrm{ac},0}=2.4\hbar\Omega$ in (b) and at $qV_\mathrm{R}^{\mathrm{ac},0}=2.8\hbar\Omega$ in (c).}
    \label{fig:app_crossings_D0}
\end{figure}

In Fig.~\ref{fig:app_crossings_D0}, we show this for the intersection bound as function of the driving amplitude in the case of harmonic driving and energy-independent transmission, see also Fig.~\ref{fig:harmonic_D0}(c). Indeed, we show the effective distribution functions $\tilde{f}_\mathrm{R}(E)$ for two values of the driving amplitude in the vicinity of the sharp step in the intersection bound. Indeed, two additional crossing points, indicated by $\varepsilon_1,\varepsilon_2$ occur, when changing $V_\mathrm{R}^{\mathrm{ac},0}$ from $2.4\hbar\Omega$ to $2.8\hbar\Omega$.

\subsection{Crossing points and energy-dependent transmission}

\begin{figure}[b]
    \centering
\includegraphics[width=0.95\linewidth]{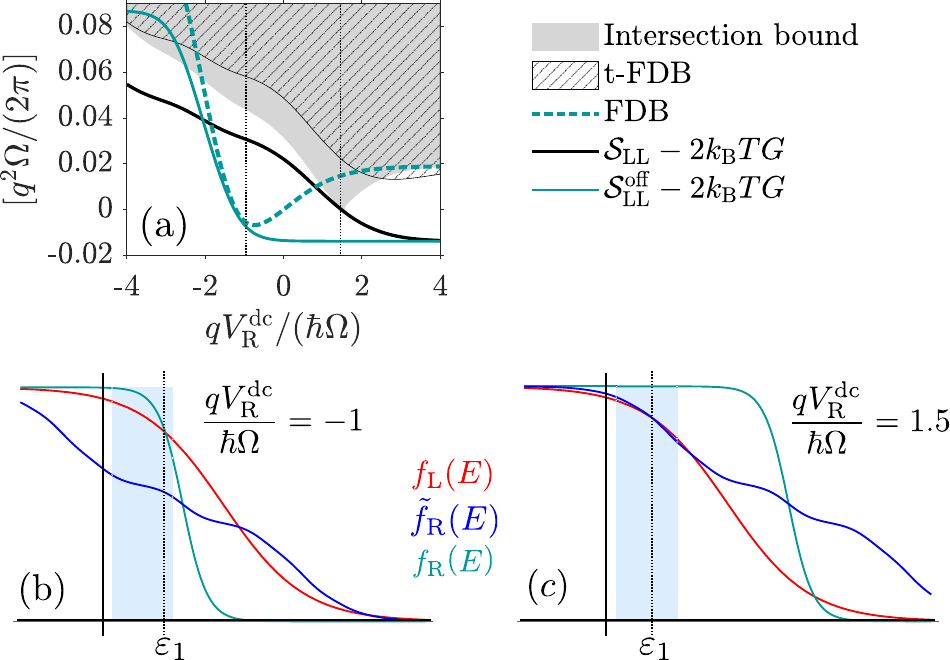}
    \caption{Excess noise and bounds for a harmonically driven two-terminal conductor with box-car shaped transmission probability $D_\mathrm{box}(E)$ [all parameters in panel (a) as in Fig.~\ref{fig:harmonic_box}(a) in the main text]. Panels (b) and (c) show the the distribution function for the hot contact (red), of the modified distribution from the cold contact due to driving (blue), and the cold contact in the absence of driving (cyan) at the parameter values indicated by vertical dashed lines in (a), namely at $qV_\mathrm{R}^{\mathrm{dc}}=-\hbar\Omega$ in (b) and at $qV_\mathrm{R}^{\mathrm{dc}}=1.5\hbar\Omega$ in (c).}
    \label{fig:app_crossings_Dbox}
\end{figure}

How tight the different bounds are and in which hierarchy they occur depends strongly on the energy-filtering properties of the transmission functions, as can be observed when comparing Fig.~\ref{fig:harmonic_D0} with \ref{fig:harmonic_box}. We show this for two different parameter sets and the related distribution function in Fig.~\ref{fig:app_crossings_Dbox}.

At $qV_\mathrm{R}^\mathrm{dc}=-\hbar\Omega$, indicated by the left vertical dashed line in Fig.~\ref{fig:app_crossings_Dbox}(a), the excess noise of the time-dependently driven system breaks the static FDB, while the t-FDB and intersection bounds remain valid. Also, the static bound has an opposite sign compared to the bounds in the presence of driving. Panel~(b) shows the relevant distribution function, where the region selected by the box-car-shaped energy filter is highlighted in light blue. Indeed, when comparing the reference hot distribution with the distribution of the right, colder contact, one notices that for the driven case the effective cold distribution $\tilde{f}_\mathrm{R}(E)$ is smaller than the hot reference distribution, while for the static case, the cold distribution $f_\mathrm{R}(E)$ is larger than the hot one. This means that their relative magnitudes are inverted within the energy window selected by the filter, when comparing the driven and static case. Also, the cold distribution of the static system (cyan) is much closer to the hot distribution than the modified cold distribution of the driven system (blue). Therefore the static and the time-dependently driven case are expected to behave fundamentally different. The static FDB is close to zero, while the t-FDB and intersection bounds are positive and larger, such that the excess noise in the driven case can break the static bound. 

Instead, at $qV_\mathrm{R}^\mathrm{dc}=1.5\hbar\Omega$, indicated by the right vertical dashed line in Fig.~\ref{fig:app_crossings_Dbox}(a), the intersection bound is close to zero. Indeed, analyzing the distribution functions at this point, shows that the energy filter exactly selects the interval around the crossing points between the effective driven distribution and the hot reference distribution, where the two distributions are furthermore very similar.

In both cases, the energy filter $D(E)$ selects a rather small window of the hot distribution, $w\ll k_\mathrm{B}T_\mathrm{L}$, such that the (hot reference) distribution assumes a close to constant value $f_\mathrm{L}\approx f_\mathrm{L}(\varepsilon_1)$ in the relevant energy interval. The bounds in both indicated situations are therefore relatively tight, see also the derivation of the bounds in Sec.~\ref{sec:FDB}.

\bibliography{refs.bib}

\end{document}